%% file: main.tex
\documentclass[letterpaper, 10 pt, conference]{ieeeconf}  
\IEEEoverridecommandlockouts                              
\usepackage{mathrsfs} 
\usepackage{kpfonts,comment} 
\usepackage{times}   
\usepackage{cancel}
\usepackage{xz_style}
\usepackage{setspace}
\usepackage{booktabs}
\usepackage{siunitx}
\usepackage{sidecap}
\usepackage{{makecell}}
\setcellgapes{3pt}\makegapedcells
\usepackage{cite}
\usepackage{arydshln}
\ADLdrawingmode{2}
\usepackage{array,multirow}

\title{\LARGE \bf Decision Transformer as a Foundation Model for \\ Partially Observable Continuous Control}

\author{Xiangyuan Zhang$^{1*}$ \quad Weichao Mao$^{1*}$ \quad Haoran Qiu$^{2}$ \quad Tamer Ba\c{s}ar$^{1}$
\thanks{$^*$Equal contributions. $^{1}$Department of ECE and CSL, University of Illinois Urbana-Champaign (UIUC). $^{2}$Department of CS and CSL, UIUC. Emails: \{xz7, weichao2, haoranq4, basar1\}@illinois.edu.}
}

\begin{document}

\maketitle
\thispagestyle{empty}
\pagestyle{empty}

\begin{abstract}
Closed-loop control of nonlinear dynamical systems with partial-state observability demands expert knowledge of a diverse, less standardized set of theoretical tools. Moreover, it requires a delicate integration of controller and estimator designs to achieve the desired system behavior. To establish a general controller synthesis framework, we explore the Decision Transformer (DT) architecture. Specifically, we first frame the control task as predicting the current optimal action based on past observations, actions, and rewards, eliminating the need for a separate estimator design. Then, we leverage the pre-trained language models, i.e., the Generative Pre-trained Transformer (GPT) series, to initialize DT and subsequently train it for control tasks using low-rank adaptation (LoRA). Our comprehensive experiments across five distinct control tasks, ranging from maneuvering aerospace systems to controlling partial differential equations (PDEs), demonstrate DT's capability to capture the parameter-agnostic structures intrinsic to control tasks. DT exhibits remarkable zero-shot generalization abilities for completely new tasks and rapidly surpasses expert performance levels with a minimal amount of demonstration data. These findings highlight the potential of DT as a foundational controller for general control applications.
\end{abstract}

\section{Introduction}
\input{intro}

\section{Preliminaries}
\input{setting}

\section{Learning a Single Control Task}\label{sec:single_task}
\input{single_task}

\section{Generalizing to Unseen Control Tasks}\label{sec:multi_task}
\input{multi_task}

\section{Discussion and Conclusion}\label{sec:conclusion}
\input{conclusion}

\normalsize
\section*{Acknowledgment}
XZ and TB were supported in part by the US Army Research Office (ARO) Grant 24-1-0085 and in part by the ARO MURI Grant AG285. WM, HQ, and TB were supported in part by the National Science Foundation (NSF) Grant No. CCF 20-29049 and in part by the IBM-ILLINOIS Discovery Accelerator Institute (IIDAI).


\bibliographystyle{unsrt_abbrv_custom}   
\bibliography{main}

\appendix
\counterwithin{figure}{subsection}
\counterwithin{table}{subsection}
\input{appendix}

\end{document}

%% file: intro.tex
Effective closed-loop control of nonlinear systems with partial-state observability demands a diverse spectrum of theoretical tools from adaptive and robust control to state estimation \cite{anderson1979optimal, bacsar1995h, zhou1996robust, aastrom2008adaptive}. In addition, expert-level domain knowledge is required to carefully integrate different components of the control system to achieve the desired behavior in each application. To establish an end-to-end controller synthesis framework that is agnostic to system dynamics, control objectives, and policy constraints, Reinforcement Learning (RL) has taken the initial step \cite{schulman2015high,lillicrap2015continuous,recht2019tour,hu2023toward}. However, as RL methods focus on optimizing a single task, RL controllers can hardly generalize to new tasks with different system parameters, an updated objective, or a set of new constraints. Upon each task change, RL requires millions of new samples for re-training, hindering its applications to systems with rapidly changing dynamics or industries that require an adaptive controller, e.g., aerospace \cite{lavretsky2012robust}. Recently, meta-RL has been proposed for solving these multi-task scenarios, where it learns a policy for handling multiple tasks simultaneously, and adapts to unseen tasks using minimal iterations or demonstrations \cite{finn2017model, finn2019online, mitchell2021offline}. Applying meta-RL to adaptive control and investigating its theoretical properties have been a trending research topic \cite{shi2021meta, richards2023control, musavi2023convergence, toso2024meta}. 

Inspired by the success of Large Language Models (LLMs) \cite{radford2019language, brown2020language, achiam2023gpt, touvron2023llama}, we explore the Transformer architecture \cite{vaswani2017attention} for learning controller designs, building on the Decision Transformer (DT) literature \cite{chen2021decision, janner2021offline}. First, we formulate the control problem as predicting the current optimal action from a sequence of past observations, actions, and rewards. This approach leverages Transformer's exceptional representational capabilities and the inherent structure of control tasks, effectively compressing the history into an ``approximate information state'' \cite{subramanian2022approximate} and eliminating the explicit state estimation step \cite{astrom1971introduction, anderson1979optimal, zhang2023learning, zhang2023global}. 

We initialize DT using a pre-trained GPT-2 model with $117$ millions of parameters for language tasks \cite{radford2019language}, and train DT on a self-generated offline control dataset via Low-Rank Adaptation (LoRA) \cite{hu2021lora}. Then, we evaluate DT's effectiveness through comprehensive experiments on five distinct tasks, from maneuvering aerospace systems to controlling partial differential equations (PDEs). In Section \ref{sec:single_task}, we first study the single-task control setting as a proof of concept, showing that DT can consistently match or surpass the unknown behavior policies and is also competitive with state-of-the-art offline RL methods \cite{kumar2020conservative}. We detail a few single-task ablation studies in Section \ref{sec:single_ablations}. Subsequently, Section \ref{sec:multi_task} investigates the multi-task setting with significantly perturbed system dynamics, either through altering physical parameters or injecting random noises into the system matrices. We show that DT can zero-shot generalize to new tasks that are completely out of the training distribution. Moreover, after seeing a minimal amount of demonstration data (e.g., $10$ rollout trajectories), DT can quickly match or outperform expert behavior policy in the new task. Our experimental findings confirm DT's ability to capture parameter-agnostic structures of control tasks and excel in few-shot learning, suggesting its potential for general control applications as a foundational controller. In Section \ref{sec:conclusion}, we discuss various design choices and list several future research directions.

\subsection{Related Literature}
RL methods have surpassed human performance in many decision-making tasks with full-state observability \cite{mnih2015human, lillicrap2015continuous, silver2017mastering}. However, real-world control tasks often involve noisy, lower-dimensional, and delayed observations, requiring policies to infer or reconstruct missing state information \cite{astrom1971introduction}. Control tasks also require policies to stabilize the closed-loop system over an extended period, in addition to maneuvering the intense dynamics of aerospace systems \cite{lavretsky2012robust} or turbulent flows \cite{brunton2015closed} to avoid catastrophic failures. Lastly, the scalability issue arises in high-dimensional control problems. Motivated by these unique challenges, controlgym was introduced in \cite{zhang2023controlgym} to benchmark RL-based controllers. Another line of research also studies the theoretical foundation of RL for learning control policies \cite{recht2019tour, fazel2018global, zhang2019policymixed, zhang2021derivative, zhang2023revisiting, hu2023toward}.

Motivated by LLMs' success, the Transformer architecture has been widely applied and evaluated in fully observable decision-making and control tasks \cite{chen2021decision, janner2021offline, lee2022multi, xu2022prompting, zheng2022online, yamagata2023q, xu2023hyper, shi2023unleashing, schmied2024learning}, particularly on D4RL \cite{fu2020d4rl},  Meta-World \cite{yu2020meta}, and DM\_Control \cite{tassa2018deepmind} benchmarks. Specifically, \cite{chen2021decision} introduced DT for modeling RL as a sequence prediction problem. \cite{lee2022multi} applied DT to learn a suite of Atari games simultaneously and achieved human-level performances. \cite{xu2022prompting, xu2023hyper} generalized DT to new tasks by prompting demonstration trajectories and adding task-specific adaptation modules, respectively. \cite{yamagata2023q} improved learning from sub-optimal demonstration trajectories by incorporating Q-values into DT's return-to-go input modality. \cite{zheng2022online} introduced an online fine-tuning step to improve learning from sub-optimal demonstrators. \cite{shi2023unleashing} refined the DT architecture and proposed co-training of language and decision-making tasks. \cite{schmied2024learning} investigated generalizations to new tasks without forgetting the previously learned skills. Besides the DT literature, \cite{esslinger2022deep} leveraged a Transformer architecture to predict Q-values in a Deep-Q-Network algorithm for partially observable RL problems. In contrast, our work builds upon the DT literature, especially the most recent developments in \cite{shi2023unleashing}, to directly predict actions from history in partially observable control.

In parallel to our investigations, a significant body of literature have leveraged Transformer models to auto-regressively generate control inputs in robotic systems and embodied agents \cite{reed2022generalist, brohan2022rt, jiang2022vima, gupta2022metamorph, brohan2023rt, bousmalis2023robocat, zhao2023learning, team2023octo, radosavovic2023robot, radosavovic2024humanoid}. For comprehensive reviews on this topic, we refer the readers to \cite{firoozi2023foundation, kawaharazuka2024real}. These studies have focused on tokenizing multimodal inputs, including vision, language, sensor data, and robotic actions, and utilizing the Transformer backbone to train visual-language-sensory-motor policies. The overarching goal is to develop robotic foundation models that integrate traditional perception, planning, and control stacks into a unified framework. In contrast, our work provides a control-centric perspective, aiming to thoroughly explore both the potential and the challenges of applying Transformers solely to control problems.

%% file: setting.tex
\begin{figure*}[!ht]
    \centering
    \includegraphics[width=0.95\textwidth]{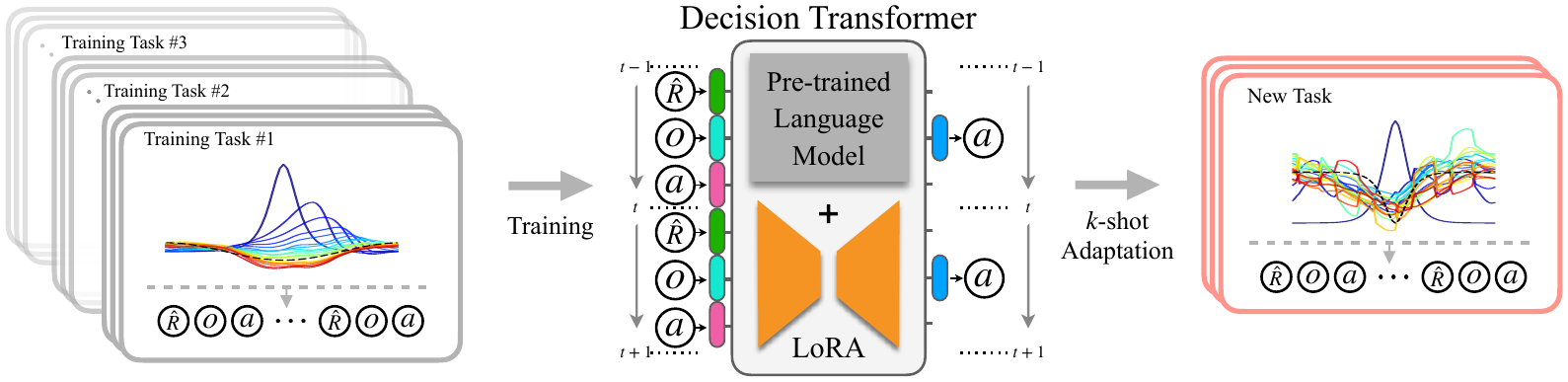}
    \vspace{-0.5em}
    \caption{Illustration of the DT Architecture. \textit{Left:} An offline control dataset sampled from some unknown behavior policies. \textit{Middle}: DT predicts an optimal action $a_t$ autoregressively based on an input sequence of `reward-to-go's, observations and actions. The pre-trained language weights of DT are kept frozen while we employ LoRA for control training. \textit{Right}: DT quickly generalizes to new control tasks after seeing minimal offline demonstrations.}
    \label{fig:DT}
    \vspace{-1.2em}
\end{figure*}

We consider the discrete-time nonlinear system dynamics
\begin{align}\label{eqn:nonlinear_dynamics}
    s_{t+1} = f(s_t, a_t; w_t), \quad o_{t} = g(s_t; v_t),
\end{align}
where $f:\RR^{n_s} \times \RR^{n_a} \to \RR^{n_s}$ and $g:\RR^{n_s} \to \RR^{n_o}$ are time-invariant mappings that define the state transition and observation process, respectively. We use  $s_t$ to denote the state, $a_t$ to denote the action/control input, $o_t$ to denote the observation, and $w_t, v_t$ to denote noises and disturbances. We define the one-step reward to be the negative quadratic tracking cost

\vspace{-1em}
\small
\begin{align*}
    r_t(s_t, a_t) = -(s_t-s_r)^{\top}Q_1(s_t-s_r) - a_t^{\top}Q_2a_t - 2(s_t-s_r)^{\top}Q_3a_t,
\end{align*}
\normalsize
where $Q_1 \geq 0$ and $Q_2 > 0$ are symmetric weighting matrices, $Q_3$ is another weighting matrix in $\RR^{n_s \times n_a}$, and $s_r \in \RR^{n_s}$ is the target state. We additionally let $r_t := r_t(s_t, a_t)$. The objective is to compute a policy $\pi$ that maps the available information up to and including step $t$, i.e., $\{o_0, a_0, r_0, \cdots, o_t\}$, to the current action $a_t$, so as to maximize the expected cumulative reward $\EE \sum_{t=0}^{T}\gamma^t r_t$. Here, $\gamma \in [0, 1)$ is the discount factor and $T$ is a possibly infinite problem horizon length\footnote{Our setting can be equivalently viewed as a partially observable Markov decision process (POMDP) \cite{kaelbling1998planning}, denoted by the tuple $\langle S, A, T, R, \Omega, O, \gamma \rangle$. The $S$, $A$, and $\Omega$ in POMDP correspond to the continuous spaces $\RR^{n_s}$, $\RR^{n_a}$, and $\RR^{n_o}$ in our setting. The reward function $R$ corresponds to our $r_t$. The transition and observation models $T$ and $O$ correspond to $f$ and $g$ in (\ref{eqn:nonlinear_dynamics}).}.

\subsection{Learning a Single Control Task}
In decision-making and control tasks, model-free RL can be generally categorized into online and offline methods. Online RL directly interacts with the environment, utilizing real-time feedback to refine the control policy. Conversely, offline RL learns a control policy from a pre-collected dataset of system trajectories that are sampled using some unknown behavior policies \cite{levine2020offline}. In this work, we study the offline RL setting for control. Suppose we have access to $\cD$, a set of system trajectories generated by some unknown behavior policies. Each trajectory has the form of
\begin{align}\label{eqn:dataset}
\{o_0, a_0, r_0, \cdots, o_T, a_T, r_T \}, \quad T \in \ZZ^+.
\end{align}
The objective is to learn a control policy solely from the dataset $\cD$ and achieve a matching or higher cumulative reward than the unknown behavior policy. 

Compared with standard offline RL \cite{levine2020offline, kumar2020conservative}, Transformer's superior representation ability enables compressing history into an ``approximate information state'', serving as a proxy for predicting the true system evolution \cite{subramanian2022approximate}.

\subsection{Control of Multiple Systems and Meta-RL}
In simultaneously controlling multiple systems or a system with varying parameters, the controller must quickly adapt to unseen scenarios. Adaptive control methods are widely used for this purpose \cite{aastrom2008adaptive}, notably in aerospace control systems \cite{lavretsky2012robust}. Alternatively, $H_{\infty}$ control concerns a continuum of systems (slightly) perturbed from the nominal dynamics, enabling a controller based on the nominal dynamics to maintain performance in the perturbed environment without modification, i.e., a zero-shot approach \cite{bacsar1995h, zhou1996robust}. However, not having a learning component, $H_{\infty}$ control design cannot cope with substantial perturbations and may compromise performance excessively for robustness.

Recently, meta-RL has been proposed to train on multiple tasks and then adapt rapidly to unseen tasks with minimal iterations or demonstrations, respectively, in online and offline contexts \cite{finn2017model, finn2019online, mitchell2021offline}. Its application to continuous control tasks has been studied in \cite{shi2021meta, richards2023control,musavi2023convergence,toso2024meta}. In the offline setting, the goal of meta-RL is to learn a policy from $\cD$ that can act as an effective initial point for rapid adaptation to similar new tasks drawn from the same distribution. 

\subsection{Decision Transformer as a Foundation Model for Control}
DT models control tasks as sequence prediction problems \cite{chen2021decision}. At time $t$, DT predicts an optimal action $a_t$ autoregressively based on history $\{\hat{R}_{t-K+1}, o_{t-K+1}, a_{t-K+1}, \cdots, \hat{R}_{t}, o_t\}$ of context length $K$. Here, $\hat{R}_t:=\sum_{\tau=t}^{T}r_{\tau}$ represents the reward-to-go. We utilize three multi-layer perceptrons (MLPs) for embedding raw inputs (i.e., rewards-to-go, observations, and actions) alongside an additional MLP for generating actions \cite{shi2023unleashing}, as depicted in Figure \ref{fig:DT}. 

We adopt the pre-trained GPT-2 language model with $117$ millions of parameters \cite{radford2019language} to initialize DT. Subsequently, we train DT on the control dataset $\cD$ using LoRA \cite{hu2021lora}. LoRA keeps the pre-trained language weights frozen and adjusts DT by training rank decomposition matrices, reducing the number of trainable parameters by orders of magnitudes. 

After training, we can apply DT to new, unseen control tasks in a zero-shot approach. Alternatively, we can further adapt DT using a few demonstration trajectories of the new task. Hereafter, we refer to this approach as $k$-shot adaptation. The required number of demonstrations, $k$, is typically much lower than that of offline RL from scratch thanks to the pre-trained language weights and control training.

In the inference phase, we fix the DT weights and use them to directly interact with the control environment. Unlike the training stage, access to the reward-to-go from a dataset is not available. Consequently, we substitute DT's input modality $\hat{R}_{t}$ with a prompt of the desired target return, i.e., the discounted cumulative rewards \cite{chen2021decision}. This target return may be the maximal achievable return for this task or a specific performance threshold. Based on the desired target return and observations, DT predicts an action to be executed by the actuators. Then, upon receiving the one-stage reward, we decrease the desired target return by the same amount and use it as the next prompt input. This process is repeated until the end of the episode.

%% file: single_task.tex

As a proof of concept, we first train DT using LoRA to learn a single control task from offline demonstrations. Our experiments include five controlgym tasks \cite{zhang2023controlgym}: the helicopter and aircraft examples (he1) and (ac4), a high-dimensional cable-mass system (cm3), and PDE control problems (CDR) and (Burgers). The detailed settings are listed in Appendix \ref{sec:setup}.

In each task, we generate the ``expert dataset'' using the converged Proximal Policy Optimization (PPO) policy \cite{schulman2017proximal} while sampling the ``medium dataset'' using an early-stopped PPO policy. We then normalize the cumulative rewards to the range of $[0, 1]$, where the expert behavior scores $1$ and the medium behavior scores $0$. We refer to Section \ref{sec:single_ablations} for ablation studies in dataset generation.

\subsection{Single-Task Experimental Results}

\renewcommand{\arraystretch}{1.15}
\begin{table*}[t]
\centering\caption{Learning a Single Control Task from Offline Demonstrations}\label{table:single_task}
\vspace{-0.6em}
\footnotesize
\begin{tabular}[width=0.95\textwidth]{|c|c:c:cccc|}
\hline
\textbf{Dataset} & \textbf{Task}  & \textbf{PPO} & \textbf{CQL} & \textbf{BC}  & \textbf{DT} & \textbf{DT w/ co-training} \\ \hline\hline
& he1 & $1.000$  & $1.013\pm 0.074$ & $1.025\pm 0.060$ & $\mathbf{1.044\pm 0.040}$ & $1.042\pm 0.036$\\
& ac4 & $1.000$ & $1.209\pm 0.240$ & $0.692\pm 0.695$ & $\mathbf{1.410\pm 0.086}$ & $1.275\pm 0.087$\\
\textbf{Expert} & cm3 & $1.000$ & $1.146\pm 0.148$ & $0.895\pm 0.290$ & $\mathbf{1.243\pm 0.235}$ & $0.997\pm 0.224$\\
 & CDR & $1.000$ & $\mathbf{1.017 \pm 0.042}$ & $-0.016 \pm 0.571$ & $0.990 \pm 0.018$ & $0.997 \pm 0.017$\\
& Burgers  & $1.000$ & $1.171\pm 0.220$ & $1.070\pm 0.312$ & $1.210\pm 0.167$ & $\mathbf{1.319\pm 0.076}$ \\
\hline
& he1 & $1.000$ & $0.339\pm 0.297$ & $-0.724\pm 0.348$ & $0.527\pm 0.143$ & $\mathbf{0.590\pm 0.124}$\\
 & ac4 & $1.000$ & $0.127\pm 0.240$ & $-0.309\pm 0.677$ & $0.547\pm 0.210$ & $\mathbf{0.583\pm 0.224}$\\
\textbf{Medium} & cm3 & $1.000$ & $0.050\pm 0.334$ & $0.376\pm 0.483$ & $0.499\pm 0.189$ & $\mathbf{0.609\pm 0.203}$\\
& CDR & $1.000$ & $\mathbf{0.037 \pm 0.078}$ & $-0.440 \pm 0.364$ & $-0.049\pm 0.097$ & $-0.026 \pm 0.079$\\
& Burgers  & $1.000$ & $0.225\pm 0.178$ & $0.234\pm 0.171$ & $0.371\pm 0.086$ & $\mathbf{0.501\pm 0.151}$\\
\hline
\end{tabular}
\vspace{0.7em}

\small
\centering Normalized rewards in five control tasks, where we compare CQL, BC, DT, and DT with co-training.
\vspace{-2em}
\end{table*}
\normalsize

We evaluate DT's performance and compare DT against the unknown PPO behavior policies, Conservative Q-Learning (CQL), and Behavior Cloning (BC). Training on expert datasets, Table \ref{table:single_task} indicates that DT consistently matches or surpasses expert demonstrators whose scores are normalized as $1$. Furthermore, DT matches or exceeds the performance of state-of-the-art baselines across five experiments. Notably, in the aircraft example ac4, DT significantly outperforms both the demonstrator and other baselines, highlighting its capability in maneuvering the intense dynamics of aircraft flight control.

When trained on medium datasets (whose behavior policy score is normalized to $0$), the performance advantage of DT and its variant with language co-training over demonstrators and other baselines becomes even more substantial. Specifically, DT consistently achieves normalized scores between $0.5$ and $0.6$ in four out of the five experiments (he1, ac4, cm3, and Burgers). DT matches the unknown early-stopped PPO behavior policy in CDR. In contrast, the scores for CQL and BC vary widely across different experiments, and BC could fail to match medium demonstrators in the he1, ac4, and CDR environments. 

\subsection{Ablation Studies}\label{sec:single_ablations}
We conduct ablation studies on dataset generation and training DT, and list here some interesting observations.

\textbf{Non-Stationary Behavior Policy:} The optimal policy for finite-horizon control problems, such as our $50$-step linear ones, is non-stationary \cite{bacsar1995h}. Therefore, it is logical to use, e.g., the finite-horizon non-stationary linear-quadratic Gaussian (LQG) policy, to sample the expert dataset for these linear control tasks. However, DT randomly segments trajectories in training without encoding their absolute time steps, which reveals (to the policy) only the relative temporal information within each segment. This issue also arises in the fully observable (state-feedback) setting. We conjecture that adapting the DT architecture is needed for effectively learning from datasets generated by dynamic LQG control policies with an intricate in-loop state estimator. 

\textbf{On the Context Length $K$:} To avoid the above non-stationarity issue in the $50$-step linear control tasks, one can apply the infinite-horizon LQG policy, which is sub-optimal but very cheap to compute. In comparison, the PPO policy we used in the expert dataset, which is much more expensive to obtain, is also slightly sub-optimal. However, the infinite-horizon LQG policy is not a desired demonstrator to DT since the actions it generates implicitly depend on all previous observations through the evolution of a state estimator. Since the Transformer policy is high-dimensional, it is capable of fitting demonstration trajectories generated by any lower-dimensional stationary behavior policies, if their actions depend only on history with a context length of $K$ or less. For instance, PPO actions rely on immediate observations (context length of $0$), whereas LQG actions could depend on a much longer history sequence $T >> K$. This is not an issue in the fully observable Markovian setting, because current-state-feedback stationary policies such as infinite-horizon LQR are good demonstrators. 

\textbf{Language Co-Training:} We study the effects of adding an auxiliary language prediction loss to DT in our control tasks, which has been advocated in \cite{shi2023unleashing}. Table \ref{table:single_task} reports mixed outcomes from this additional language co-training. It generally fails to improve, and may even degrade, DT's efficacy when training on expert datasets, possibly because the auxiliary language prediction loss distracts DT from its control objective. In contrast, on medium datasets, co-training can substantially enhance DT's control performances, as a by-product of its original motivation of preserving the pre-trained language capabilities  (cf., Section 5.5 of \cite{shi2023unleashing}). Based on this observation and that we only train DT on expert datasets in multi-task experiments, hereafter we will primarily focus on analyzing DT without co-training.

%% file: multi_task.tex
We examine DT's ability to generalize across new control tasks. In this multi-task framework, we construct the training tasks by randomly perturbing the nominal system matrices in the single-task he1, ac4, and cm3 experiments, and randomly sample physical parameters of the CDR and Burgers tasks. Then, we use the converged PPO policy to generate the dataset. For the evaluation phase, we differentiate between in-distribution and out-of-distribution tests. In-distribution tests adhere to the training dataset's task distribution, whereas out-of-distribution tests involve either introducing larger perturbations or selecting physical parameters that deviate significantly from the training task distribution. We evaluate two versions of DT in both test scenarios: zero-shot and $10$-shot, where ``shot'' refers to the number of demonstration trajectories used for adaptation. Details on the configuration of these multi-task experiments can be found in Appendix \ref{sec:setup}. We provide two additional experiments in Subsection \ref{sec:multi_ablations}.

\renewcommand{\arraystretch}{1.15}
\begin{table}[!ht]
\setlength{\tabcolsep}{4pt}
\centering\caption{Few-Shot Generalizations to New Tasks}\label{table:multi_task}
\vspace{-0.5em}
\footnotesize
\begin{tabular}[width=0.49\textwidth]{|c|c:ccc|}
\hline
 & \textbf{Task} & \textbf{Prompt DT} & \textbf{DT (0-shot)}  & \textbf{DT (10-shot)} \\ \hline\hline
& he1 & $-0.384\pm 0.077$ & $0.860\pm 0.107$ & $\mathbf{0.979\pm 0.134}$ \\
& ac4 & $\mathbf{1.012\pm 0.063}$ & $0.679\pm 0.228$ & $0.651\pm 0.140$ \\
\textbf{In.} & cm3 & $1.151\pm 0.176$ & $1.120\pm 0.067$ & $\mathbf{1.255\pm 0.082}$ \\
& CDR & $\mathbf{1.144 \pm 0.058}$ & $0.403\pm 0.161$ & $1.013 \pm 0.004$\\
& Burgers & $1.154\pm 0.071$ & $0.663\pm 0.106$ & $\mathbf{1.203\pm 0.049}$ \\
\hline
& he1 & $-0.010\pm 0.013$ & $0.728\pm 0.137$ & $\mathbf{1.017\pm 0.159}$ \\
& ac4 &  $-0.585\pm 0.028$ & $0.626\pm 0.146$ & $\mathbf{1.129\pm 0.162}$ \\
\textbf{Out.} & cm3 & $-1.000\pm 0.000$ & $1.270\pm 0.338$ & $\mathbf{1.455\pm 0.347}$ \\
& CDR & $-0.466 \pm 0.050$ & $0.597 \pm 0.239$ & $\mathbf{1.010 \pm 0.006}$ \\
& Burgers & $0.168\pm 0.129$ & $0.626\pm 0.298$ & $\mathbf{1.071\pm 0.209}$ \\
\hline
\end{tabular}
\vspace{0.7em}
\small

\centering Normalized rewards in new tasks, averaging over $9$ tests for in-distribution (In.) and out-of-distribution (Out.) parameters. The normalization process is identical to that of the single-task experiments: the range is $[0, 1]$ with the expert behavior policy scores $1$ and the medium behavior policy scores $0$.
\vspace{-1.5em}
\end{table}
\normalsize

\subsection{Multi-Task Experimental Results}
We train DT on $30$ control tasks, and evaluate its generalization capabilities to $9$ in-distribution tests and $9$ out-of-distribution tests; see Appendix \ref{sec:setup} for details. The results of the multi-task experiments are listed in Table \ref{table:multi_task}, where we compare the normalized scores of zero- and $10$-shot DT with those of Prompt-DT (PDT) \cite{xu2022prompting}. An immediate observation from Table \ref{table:multi_task} is that the performance of PDT could deteriorate significantly in out-of-distribution tests. In contrast, DT ($10$-shot) consistently yields a strong performance in all tests. This observation demonstrates DT's rapid adaptation ability to surpass ``expert'' performance after seeing a minimal $10$ demonstrations. Moreover, DT zero-shot achieves reasonable performance across different tasks, which indicates DT's capability to capture the parameter-agnostic structure of the underlying control tasks from the training set. Our experimental results demonstrate the power of the Transformer architecture in zero-shot generalization and few-shot adaptation to new tasks.

\subsection{Additional Experiments}\label{sec:multi_ablations}
\textbf{Cross-Environment Generalization:} As a preliminary study, we generalize the DT trained on the Burgers dataset to the CDR tasks, facilitated by the shared dimensionalities of the two environments. As shown in Table \ref{table:cross_env}, DT zero-shot is ineffective since it has not seen any CDR trajectories from training or adaptation. However, with as few as 10 demonstration trajectories, DT quickly adapts to CDR tasks, achieving cumulative rewards comparable to those of the DT trained exclusively on CDR (cf., Table \ref{table:multi_task}). This preliminary study indicates that DT might be capable of generalizing to different environments in the few-shot setting.

\small
\setlength{\tabcolsep}{5pt}
\renewcommand{\arraystretch}{1.15}
\begin{table}[t]
\centering\caption{Cross-Environment Generalization}\label{table:cross_env}
\vspace{-0.6em}
\footnotesize
\begin{tabular}[width=0.45\textwidth]{|c|c:cc|}
\hline
 & \textbf{Task}  & \textbf{Burgers-DT (0-shot)} & \textbf{Burgers-DT (10-shot)} \\ \hline\hline
\textbf{In.}& CDR & $-1.000\pm 0.000$ & $1.011\pm 0.007$  \\
\textbf{Out.}& CDR & $-0.715\pm 0.016$ & $1.006 \pm 0.004$ \\
\hline
\end{tabular}
\vspace{-2em}
\end{table}
\normalsize

\textbf{Comparing to Model-Based $H_2/H_{\infty}$ Control:} We compare DT's performance against that of the output-feedback $H_2/H_{\infty}$ controller in three linear control tasks he1, ac4, and cm3. The $H_2/H_{\infty}$ controller is computed based on the unperturbed system model used in the single task experiments, as detailed in Chapter 6 of \cite{bacsar1995h}. We select the robustness parameter of the $H_2/H_{\infty}$ controller from the set $\{0.5, 1, 2, 3, 4, 5, 10, 20, 50, 100, 200, 500, 1000\}$ that yields the highest average score in in-distribution tests. This optimized $H_2/H_{\infty}$ controller is then applied to out-of-distribution tests. We note that the $H_2/H_{\infty}$ controller has access to the entire history, while DT is limited to accessing the $K$-step history. To mitigate the impact of some deficient cases, scores of the $H_2/H_{\infty}$ controller are capped within the range of $[-1, 2]$ before taking averages. Our comparison favors the $H_2/H_{\infty}$ controller in every possible aspect.

\small
\setlength{\tabcolsep}{5pt}
\renewcommand{\arraystretch}{1.15}
\begin{table}[!ht]
\centering\caption{Comparison with Model-Based H$_2$/H$_{\infty}$ Controller}\label{table:ablation_hinf}
\vspace{-0.6em}
\footnotesize
\begin{tabular}[width=0.45\textwidth]{|c|c:ccc|}
\hline
 & \textbf{Task}  & \textbf{H$_2$/H$_{\infty}$ (0-shot)} & \textbf{DT (0-shot)} & \textbf{DT (10-shot)} \\ \hline\hline
\textbf{In.}& he1 & $\mathbf{1.472 \pm 0.358}$ & $0.860\pm 0.107$ & $0.979\pm 0.134$ \\
\textbf{Out.}& he1 & $0.665 \pm 1.219$ & $0.728\pm 0.137$ & $\mathbf{1.017\pm 0.159}$ \\
\hline
\textbf{In.}& ac4 &  $-0.552 \pm 0.738$ & $\mathbf{0.679\pm 0.228}$ & $0.651\pm 0.140$ \\
\textbf{Out.} & ac4 & $-0.297 \pm 1.232$ & $0.626\pm 0.146$ & $\mathbf{1.129\pm 0.162}$ \\
\hline
\textbf{In.} & cm3 & $\mathbf{1.310 \pm 0.279}$ & $1.120\pm 0.067$ & $1.255\pm 0.082$ \\
\textbf{Out.} & cm3 & $0.522 \pm 1.359$ & $1.270\pm 0.338$ & $\mathbf{1.455\pm 0.347}$ \\
\hline
\end{tabular}
\end{table}
\normalsize

We now discuss the implications of the results in Table \ref{table:ablation_hinf}. First, as perturbation size grows, the performance of the $H_2/H_{\infty}$ controller deteriorates, aligning with our expectation from the robust control theory. Conversely, DT's performance remains unaffected or becomes even better with large perturbations in both zero-shot and $10$-shot scenarios. Therefore, we hypothesize that further growing the perturbation size would destroy the $H_2/H_{\infty}$ performance, whereas DT can continue to generalize. Moreover, in out-of-distribution tests, applying DT zero-shot achieves higher scores across all tasks than the $H_2/H_{\infty}$ controller. This observation again verifies DT's capability to capture the parameter-agnostic structure underlying each task environment.

%% file: conclusion.tex
In contrast to D4RL \cite{fu2020d4rl}, controlgym environments \cite{zhang2023controlgym} do not have a publicly available offline dataset. Therefore, we have generated our own dataset for all the experiments. We normalized the cumulative rewards to $[0, 1]$ with respect to the early-stopped and the converged PPO behavior policies, respectively, due to the varying reward scales across tasks. For the ``expert'' behavior policy normalized to $1$, we chose the PPO policy due to the absence of analytical solutions in some environments. Despite being termed ``expert'', the converged PPO policy's performance is suboptimal as it is static output-feedback, discarding the past history crucial in partially-observable decision-making \cite{bacsar1995h, subramanian2022approximate}. With access to past history, as in DT, normalized rewards greater than $1$ are achievable. For the ``medium'' behavior policy normalized to $0$, we chose the early-stopped PPO policy over a random control policy to avoid incurring near-negative infinity rewards in our control settings, which is in stark contrast to standard RL tasks. Overall, lack of a high-quality offline dataset hinders evaluating DT in other more challenging controlgym environments and extending problem horizons. This is due to the difficulties in obtaining ``good'' behavior policies. In the future, we aim to sample a high-quality offline dataset with a diverse array of behavior policies to enhance understanding of DT's efficacy.

In summary, we have investigated DT as a candidate foundational controller for general partially observable settings. By framing the control task as predicting the current optimal action based on a finite-window truncation of past observations, actions, and rewards, DT eliminates the need for a separate estimator design. Utilizing pre-trained weights from language tasks, we trained DT for control applications, demonstrating its ability to generalize to unseen tasks with minimal demonstrations effectively. Our findings open several directions for future research, including: 1) extending DT's generalization capabilities to different environments with varying state and action space dimensions; 2) adapting DT for longer or infinite problem horizons; 3) enhancing DT's inference speed to facilitate high-frequency real-time control; 4) refining DT's architecture to learn efficiently from highly non-stationary demonstration policies; and 5) embedding control-specific insights into the DT architecture.

%% file: appendix.tex
\subsection{Experiment Settings}\label{sec:setup}
Our experiments include five control tasks from the controlgym library \cite{zhang2023controlgym}, where their corresponding dimensions $n_s, n_a$, and $n_o$ and horizon lengths are listed below.

\renewcommand{\arraystretch}{1.15}
\begin{table}[!h]
\small
\centering
\begin{tabular}[width=0.48\textwidth]{|c:ccccc|}
\hline
\textbf{Task} & $n_s$ & $n_a$ & $n_o$ & n\_steps & sample\_time\\ \hline\hline
he1 & $4$ & $2$ & $1$ & $50$ & $0.05$\\
ac4 & $9$ & $1$ & $2$ & $50$ & $0.05$ \\
cm3 & $120$ & $1$ & $2$ & $50$ & $0.25$ \\
CDR & $64$ & $5$ & $10$ & $100$ & $0.1$\\
Burgers & $64$ & $5$ & $10$ & $100$ & $0.05$\\
\hline
\end{tabular}\label{table:setting}
\normalsize
\end{table}

We now describe the parameters of these environments. In he1, ac4, and cm3, we set the noise\_cov parameter to $0.01$, which determines the intensity of the random noise in both the system dynamics and the measurement process. In generating the offline dataset $\cD$ and also conducting the experiments on in-distribution generalizations, we first randomly sample vectors $\Delta A$ and $\Delta B_2$ of appropriate dimensions from a zero-mean Gaussian distribution. Then, we perturb the system matrices $A$ and $B_2$ in he1, ac4, and cm3 by normalizing $[\Delta A, \Delta B_2]$ to the size of $\big\{[0.05, 0.05], [0.1, 0.1], [0.1, 0.1]\big\}$, respectively. In generating out-of-distribution test cases, we set the perturbation sizes $[\Delta A, \Delta B_2]$ to $\big\{[0.15, 0.15], [0.2, 0.2], [0.15, 0.15]\big\}$ for tasks $\{$he1, ac4, cm3$\}$, respectively. 

\begin{figure}[t]
\centering
\includegraphics[width=0.485\textwidth]{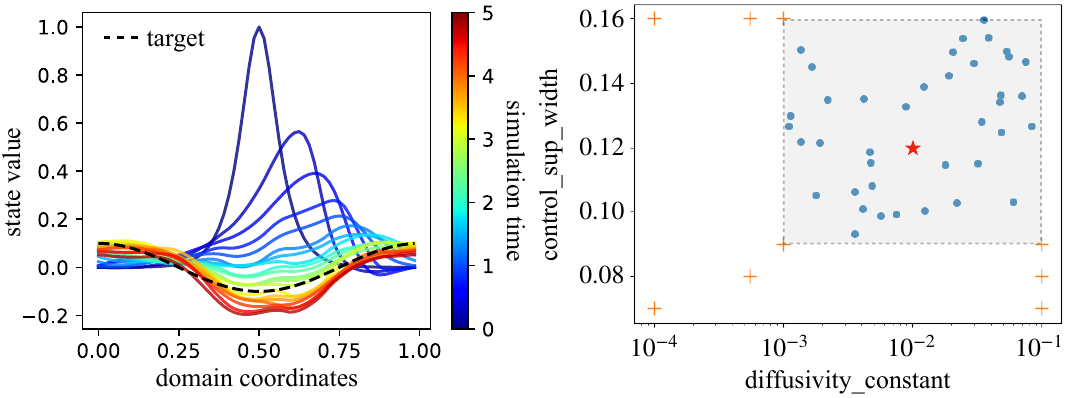}
\caption{\textit{Left:} PPO policy in Burgers with $\nu=10^{-2}$ and $\phi=0.125$. \textit{Right:} We sample the Burgers parameters uniformly from the shaded region for generating the dataset and in-distribution tests. The out-of-distribution tests are manually selected.}\label{fig:burgers}
\end{figure}

In CDR and Burgers, we perturb the system dynamics by varying the physical parameters of their governing PDE. For Burgers, we create training and in-distribution testing sets by uniformly sampling the diffusivity\_constant ($\nu$) within a logarithmic scale from $10^{-3}$ and $10^{-1}$, and also uniformly sampling the control\_sup\_width ($\phi$) between $0.09L$ and $0.16L$ (with $L=1$ denoting domain length). We deterministically generate $9$ out-of-distribution test cases as all $(\nu, \phi)$ pairs with $\nu \in \{10^{-3}, 5.5\times 10^{-4} ,10^{-4}\}$ and $\phi \in \{0.09, 0.08, 0.07\}$, presenting more challenging control tasks due to shock-like behavior from lower $\nu$ and reduced control authority with smaller $\phi$. For all cases, we set process\_noise\_cov and sensor\_noise\_cov parameters to $0$ and $0.25$, respectively. We choose the target state to the discretization of the continuous field $-0.1\cdot\cos(2\pi x / L)$ at $n_s$ equally-spaced points in the spatial domain, where $x$ represents the domain coordinate \cite{zhang2023controlgym}. Figure \ref{fig:burgers} illustrates the converged PPO policy's tracking behavior in a single Burgers experiment and the task generation process for multi-task experiments.

\begin{figure}[t]
\centering
\includegraphics[width=0.35\textwidth]{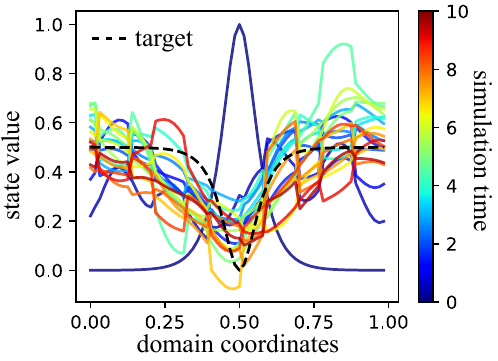}
\caption{PPO policy in CDR with $\nu=10^{-2}$, $c=0.1$, $\zeta=0$, and $\phi=0.1$.}\label{fig:cdr}
\end{figure}

Similar to Burgers, we create the CDR training and in-distribution testing sets by uniformly sampling the diffusivity\_constant ($\nu$) within a logarithmic scale from $10^{-3}$ and $10^{-1}$, the convective\_velocity ($c$) from $0$ to $0.2$, the reaction\_constant ($\zeta$) from $-0.1$ to $0.1$, and the control\_sup\_width ($\phi$) from $0.08L$ and $0.12L$ (with $L=1$). For the $9$ out-of-distribution tests, we choose $\nu, c, \zeta, \phi$ deterministically to be\par
\vspace{-1em}
\small
\begin{align*}
    &(\nu, c, \zeta, \phi) \in \Big\{(5\times 10^{-4}, 0.25, 0.15, 0.1), \\
    &\hspace{2em}(5\times 10^{-4}, 0.25, 0.2, 0.1), (5\times 10^{-4}, 0.3, 0.15, 0.1), \\
    &\hspace{2em}(5\times 10^{-4}, 0.3, 0.2, 0.1), (1\times 10^{-4}, 0.25, 0.15, 0.1), \\
    &\hspace{2em}(1\times 10^{-4}, 0.25, 0.2, 0.1), (1\times 10^{-4}, 0.3, 0.15, 0.1), \\
    &\hspace{2em}(1\times 10^{-4}, 0.3, 0.2, 0.1), (1\times 10^{-4}, 0.3, 0.2, 0.08)\Big\}.
\end{align*}
\normalsize
Compared to the training and in-distribution testing cases, these out-of-distribution parameters result in more challenging control tasks (cf., Section 3.1 of \cite{zhang2023controlgym}). For all cases, we choose the process\_noise\_cov and the sensor\_noise\_cov parameters to $0$ and $0.25$, respectively. We choose the target state to the discretization of the continuous field $0.5-0.5\cdot\mathrm{sech}(20x- 10L)$ at $n_s$ equally-spaced points in the spatial domain. Figure \ref{fig:cdr} shows the converged PPO policy's tracking performance in a CDR experiment with parameters being the mean of the training distribution.


\subsection{DT Parameters}
We have adopted the GPT-2 architecture with 117 millions of parameters\footnote{https://huggingface.co/openai-community/gpt2}. Specifically, it includes 12 Transformer layers and operates with a hidden dimension of 768. We also use 3 MLP layers, a LoRA rank of 32, and is trained with a batch size of 64. Moreover, we define a dropout rate of 0.1 to mitigate overfitting. All experiments are run on a single NVIDIA GeForce RTX 3090 GPU with 24 GB Memory.

Moreover, we employ the learning rate of $10^{-4}$, the weight decay rate of $10^{-5}$, the context length of $20$, and the `return-to-go's to be $(-50, -10)$ for he1, $(-0.3, -0.1)$ for ac4, $(-10, -5)$ for cm3, $(-1500, -300)$ for CDR, and $(-150, -110)$ for Burgers, respectively.